\documentstyle{article}

\begin{document}

\begin{center}
{\Large PHYSICS OBSERVABLES FOR COLOR TRANSPARENCY\footnote{ to
be published in the Proceedings of the Workshop: {\it Color Transparency'97},
held in Grenoble, France, June 1997}}

{\bf Bernard PIRE }

{\it Centre de Physique Th\'eorique {\footnote {unit\'e propre 014 du
CNRS}}, Ecole Polytechnique \\
F91128 Palaiseau, France}


Abstract:
The physics observables dedicated to the study of color transparency are
diverse. After a brief pedagogical introduction, we emphasize the
complementarity of the nuclear filtering and color transparency concepts. The
importance of quantum interferences leads to suspect pictures based on the
preparation of a squeezed state of small transverse size leading to a rapid
transverse expansion. The different roles of heavy and light nuclei are
emphasized. The possibility of color transparency tests in the vacuum is
stressed. Such a test may be quickly addressed in the virtual Compton scattering
reactions at CEBAF or HERMES energies.

\end{center}

Color Transparency\cite{ct} studies are entering the age of maturity. After the
first experimental data\cite{ctexp} have been analized in many details, one
understands better that different observables will shed light on this
interesting new QCD concept.

\section{ Mini-hadron scattering in a gauge field theory}
\hspace {\parindent}
Let us first present a somewhat academic but instructive calculation of the high
energy limit of a forward hadron-hadron amplitude in perturbative field
theory~\cite{cw}. We shall use the optical theorem to relate
the  total cross section to the imaginary part of the forward amplitude.

Let us first consider the two gluon exchange processes for quark-quark
scattering in the region $-t << s \sim -u $ . From the 18 diagrams which may be
drawn,  14 are vertex or self energy corrections to the one-gluon diagrams, and
as such are  real. Two of the four remaining diagrams dominate at small
$(-t)$.

 Denoting as $q$ the $4-$vector of the first
gluon emitted by the  bottom fermion line, one finds that the two diagrams have
identical boson propagators and  bottom fermion expression. From the upper
fermion line, one gets adding the two diagrams, after some simple algebra:
\begin{equation}
{\cal A} \propto {1 \over {q^- +i\epsilon}} - {1 \over {q^- -i\epsilon}}
\end{equation}
leading to a $\delta(q^-)$ constraint. Moreover the bottom fermion line
leads to a
$\delta(q^+)$ constraint, so that the $d^4q$ integration boils down to a
$2-$dimensional
transverse space integration. The resulting amplitude is dominantly imaginary:

\begin{equation}
{\cal M} = i C  s \int {{d^2q_T} \over {(2\pi)^2}} {1 \over {(q^2_T+\lambda^2)
((q_T-\Delta_T)^2+\lambda^2)}}
\end{equation}
which may be rewritten as
\begin{equation}
{\cal M} = i C  s \int d^2b e^{i\Delta_T.b} ({{g^2 K_0 (\lambda b)}\over
{2\pi}})^2
\end{equation}
where $\Delta_T$ is the (small) $2-$dimensional transfer between the
initial and
final fermion.
The important features of this result is the $2-$dimensional nature of the
integral
 and the squared factor  reminiscent of the
eikonal nature of forward amplitude in the high energy limit.

Let us now go to the almost physical case of meson-meson scattering, that is
consider  a color singlet formed by a quark -antiquark pair scattering on
another pair.
 A  straightforward computation shows that the color factors are the
same for all $2-$gluon exchange graphs connecting the upper hadron to the lower
one, but that an extra minus sign is attached for an antiquark line. The result
is then
\begin{eqnarray}
&{\cal M} = i C  s \int d^2b e^{i\Delta_T.b}\int d^2r_1d^2r_2
\psi^*(r_1)\psi^*(r_2) \\
&\{V(x_1-x_3)-V(x_2-x_3)-V(x_1-x_4)+V(x_2-x_4)\}^2 \psi(r_1)\psi(r_2)\nonumber
\end{eqnarray}
with
\begin{equation}
V(x) = {1 \over {(2\pi)^2}} \int {{d^2k e^{ik.x}} \over {k^2+ \lambda^2}}
\end{equation}
Suppose now that $r_1 = x_1-x_2$ is small (a mini-hadron); then one may
approximate
$V(x_1-x_3)-V(x_2-x_3) \sim r_1. \nabla (x_1-x_3) $ and similarly for
$V(x_1-x_4)-V(x_2-x_4)$, and get:
\begin{equation}
{\cal M} \propto r_1^2 .
\end{equation}
leading to a total cross section for the mini hadron scattering on a hadron:
\begin{equation}
\sigma \sim {Im {\cal M}\over  s} \propto r_1^2 .
\label{ct}
\end{equation}
This is color transparency. Note that if the $q \bar q$ was in a color
octet state,
 the resulting amplitude would contain an additionnal term
\begin{equation}
{\cal M} \propto  \dots + \{(V(x_1-x_3)-V(x_1-x_4)\}.\{V(x_2-x_3)-V(x_2-x_4)\}
\end{equation}
which does not vanish in the limit $x_1 \rightarrow x_2$. The color singlet
nature of the
quark-antiquark pair is thus essential for Eq.\ref{ct} to be valid.

\section {\bf How can a hadron behave like a mini-hadron?}

The understanding of color transparency experiments would be easy if we had
access to hadrons of tunable transverse size. This is almost the case for
quarkonia, whose size is governed by the inverse of the heavy quark mass. But
life is complicated in this case by the problem of the production process which
is still much debated. Hard exclusive scattering offers a promising way since it
has been recognized that these processes asymptotically select a compact valence
component of the hadron light-cone wave-function, thus forcing a hadron to
behave like a mini-hadron. The study of the (reduced) final state interactions
of these mini hadrons with spectator hadrons enable us to study the soft
scattering of these objects, opening the domain of the strong interactions of
small color-singlet objects. This should be of crucial importance for the
detailed understanding of Pomeron physics.

The selection of the compact (sometimes optimistically called point-like)
component of the hadronic wave function should not be misidentified as the
preparation (in the sense of quantum mechanics) of a squeezed  wave-packet
which would have a dramatic tendency to transversally expand at a rate governed
by Heisenberg uncertainty relations such as $\Delta p_T \Delta b_T > \hbar$. On
the contrary, one should realize that the selection process is due to a subtle
destructive interference pattern which annihilates all the contributions from
the large states. The perturbative treatment of the simplest occurence of this
phenomenon, namely the meson form factor, is much instructive. The pattern of
gluon radiation\cite{ls} indeed leads to a severe Sudakov suppression\cite{Sud}
of the amplitude when the transverse separation of quarks excedes some value of
order $(1/Q)$.

As often in quantum field theories, one should not be surprised that a reasoning
based on the trajectories of the hadronic components, or equivallently to the
expanding transverse hadronic size, leads one to erroneous statements. The
observable to be calculated, although hopefully factorizable in  short and long
distance quantities, is not adequately described by a semi-classical picture
where the process is chronologically cut into pieces: firstly, preparation of a
state through some hard scattering process, secondly, evolution of this state,
and finally soft scattering and absorption phenomena.

\section {\bf Nuclear Filtering in heavy nuclei}

\hspace {\parindent}
 The first piece of evidence for something like
color transparency came from the Brookhaven experiment on pp elastic
scattering at $90^\circ$  CM in a nuclear medium ~\cite{ref:ASC88}{}.
These data lead to a lively debate. The special feature of hadron hadron elastic
scattering at fixed angle is that in addition to a clear cut  short distance
amplitude,  there is an infrared sensitive process (the so-called
independent scattering mechanism) which allows not so small configurations to
scatter elastically. The phenomenon of colour transparency is thus
replaced by a {\it nuclear filtering} process: elastic scattering in
a nucleus filters away the big component of the nucleon wave function
and thus its contribution to the cross-section. Since the presence of these
two competing processes had been analysed \cite{ref:PR82}{} as responsible for
the oscillating pattern seen in the scaled cross-section $s^{10}d\sigma/dt$,
an oscillating color transparency ratio emerges~\cite{ref:RP90}.
One way ("attenuation method") to understand data is to define a survival
probability
 related in a standard way to an effective attenuation cross section
 $\sigma_{eff}(Q^2) $ and to plot this attenuation cross section as a
 function of the transfer of the reaction\cite{ref:JR}{}. One indeed obtains
 values of  $\sigma_{eff}(Q^2) $ decreasing with $Q^2 $ and quite smaller
 than the free space inelastic proton cross section. The survival probability
is even found\cite{ref:PR}{} to obey a simple scaling law in
$Q^2/A^{1/3}$.

The SLAC NE18 experiment\cite{ref:SLAC} measured the color transparency ratio
up to
$Q^2=7 GeV^2$ , without any observable increase. This conclusion follows
only if
the hard scattering part of the process is assumed to be the same as in
free space,
canceling out in forming the transparency ratio. While this assumption is
not needed
in the attenuation method, the precision of the data were not sufficient to
conclude much using the less model-dependent test. While the majority view
is that
these data cast doubt on the most optimistic onset of color transparency,
emphasizing
 the importance of a sufficient boost to get the small state quickly out of
the nucleus,
this conclusion remains tentative and something to be tested.

The diffractive electroproduction of vector mesons at
 Fermilab~\cite{ref:Fermilab}{} and Cern~\cite{ref:NMC}{}  exhibit
an interesting increase of the transparency ratio for data at $Q^2 \simeq 7
GeV^2$.
In this case the boost is high since the lepton energy is around
$E \simeq 200 GeV$ but the problem is to disentangle diffractive
from inelastic events.

\section{Color Transparency signals with Small Nuclei }

Although color transparency was first mostly considered as an effect to be
studied  on rather large nuclei, it became  recently clear that small nuclei
had much to teach us about this physics item. Deuteron  electrodesintegration
reactions $d(e,e'p)n$ for instance~\cite{deut}, both polarized and unpolarized,
is much interesting. The idea is  simple; let us examine the case where the
 virtual photon mostly hit the proton in the deuteron. The
   neutron momentum distribution is then  due to the combination of
both  Fermi momentum effects in the initial state
and final state interactions.
These two components are well known at small $Q^2$.
At large  $Q^2$, the hard process selects small-sized proton, and the
interaction {\it miniproton}-neutron is much weaker. This is where Pomeron
 physics enters, at least if energy is high enough. High $Q^2$
electroproduction
data appear then as an unexpected testing bench of soft physics, with the
important bonus of a controlable variable sized hadron scattering on a
normal one.
Whether the small size justifies completely a  perturbative treatment of
this small transfer amplitude is still an open question.

The cross section for  $d(e,e'p)n$ may be written as

\begin{equation}
{d\sigma \over {dE_{e'}d\Omega_{e'}d^3p_p }} = \sigma_{ep} ~~D_d(q,p_p,p_n) ~~
\delta (q_0-M_d-E_p-E_n)
\end{equation}

\noindent
where $D_d(q,p_p,p_n)$ is the joint probability of the
 initial proton having a Fermi momentum $p_p-q$ and the final proton (neutron)
a momentum $p_p$ ($p_n$). A poor energy resolution would restrict the physics
to a qualitative
 observation of color transparency, whereas a very good one would allow a
quantitative determination of the  miniproton-neutron scattering cross-section
as a  function of the miniproton size ({\it i.e.} $Q^2$), provided one
controls and
deconvoluates Fermi momentum effects in the deuteron.

Frankfurt {\it et al}~\cite{deut} estimate sizable  effects at CEBAF
energies, which amount to
$Q^2$ values in the range $\sim 4 GeV^2 \leq Q^2\leq~10~(GeV/c)^2$. Prospects
are brighter within ELFE conditions.

\section{Helicity (non-)conservation and Color Transparency signals}
 The hadronic helicity conservation rule in hard exclusive reactions~\cite{hel}
 follows from two assumptions:
\begin{itemize}
\item quark masses can be  neglected;
\item  valence states (with only fermions) dominate.

\end {itemize}
\noindent
These assumptions which asymptotically are quite solid in reactions such as
 $(e,e',p)$ are less justified in the hadronic case~\cite{gpr}. They are
exactly
what leads to the result of color transparency. It thus follows that
{\it the helicity non-conserving contributions must be  filtered away
 in a nuclear medium}.
It is thus most interesting, at a given value of  $Q^2$ to compare
the nuclear absorbtion of  amplitudes  violating the helicity conservation rule.
 For this measurement to be possible, we must consider cases where such
amplitudes
are quite large at reasonable $Q^2$ values. This is not the case
for the proton form factor $F_2$, but maybe for the $ p-\Delta$ transition form
factor~\cite {Burkert}, measured at $Q^2 = 3.2 GeV^2$.

In the case of hadronic reactions, it has been predicted~\cite{gpr} that
the amount
of helicity non conservation seen for instance in the helicity matrix
elements of the  $\rho$ meson produced in
$\pi p\rightarrow\rho p$ at 90$^\circ$ would be filtered out in a nucleus.
 Experimental data in free space~\cite{hep} yield
$\rho_{1-1}=0.32\pm0.10$, at $s=20.8$GeV$^2$, $\theta_{\rm CM}=90^\circ$,
for the non-diagonal helicity violating matrix element. If the persistence of
helicity non-conservation is correctly understood as due to independent
scattering processes which do not select mini-hadrons and thus are not
subject to color transparency, helicity conservation should be restored
at the same $Q^2$ in proceeses filtered by  nuclei. One should thus observe
a monotonic decrease of $\rho_{1-1}$ with $A$.

\section{ Color Transparency in the vacuum, the virtues of virtual Compton
Scattering}

Since the vacuum is highly non-trivial in QCD, one may dream  of
observable color transparency effects without any nuclear
reinteraction. Let us illustrate this statement via the study of
synchrotron radiation due to the bending of quarks moving in the vacuum.

 Synchrotron radiation of soft photons has been
proposed\cite{BHN} as a powerful way to test the vacuum structure of QCD. The
physical picture\cite{NR} is the following: consider the QCD vacuum as described
by domains of size $a$ (phenomenologically determined around 0.35 fm) where
colour fields are highly correlated. A fast quark propagating in this domain
will be subject to a Lorentz chromomagnetic force which will bend its trajectory
and thus force it to emit synchrotron radiation.  Thus, for a hadron of normal
transverse size of order 1 fm, one is allowed to add the contributions to
synchrotron radiation from the partons in a hadron incoherently. The story will
be completely different for a mini-hadron of  tunable size as expected in hard
exclusive reactions. Asymptotically, the part of the wave function which is able
to contribute to the scattering amplitude is of zero transverse size and the
color singlet nature forbids it to interact with the chromo-magnetic background
field. A vanishing Lorentz force will then lead to a vanishingly low
synchrotron radiation rate.

Consider now the exclusive electroproduction of a photon on a proton:

\begin{equation}  \label{ep}
  e(k) + p(p) \rightarrow e(k') + p(p') + \gamma (q')
\end{equation}
where the Bethe Heitler process and the virtual Compton scattering
amplitudes interfere.  This reaction has been much discussed recently in a
particular kinematical range as a tool to extract off-diagonal partonic
distributions\cite{VCS}. The other kinematical range we may focus is the very
soft photon case. The transverse size of the incoming and out-going hadrons
is a
crucial parameter in the description of a QCD-synchrotron radiation process
since
the gauge nature of QCD leads to a destructive interference
pattern when one coherently adds contributions of the different partons
constituting a colorless object, leading eventually to a vanishing radiation
amplitude at zero transverse separation.

 To quantify the expected effect, we need a model which interpolates between
incoherent synchrotron radiation for a low $Q^2$ process to a coherent sum of
the three valence quark radiation amplitude at intermediate $Q^2$ and a
vanishing
rate at large $Q^2$. Work is in progress along these lines\cite{dgp}.

 Let us finally note that the neutron case
$ e + n \rightarrow e + n + \gamma $
may be particularly intersting since it clearly suppresses much
contamination from usual bremsstrahlung processes.

\section {Conclusion}

  The $15-30 GeV$ continuous electron beam ELFE project is
presently discussed at the European level~\cite{ref:ELFE}. Besides the
determination of hadronic valence wave functions through the careful study of
many exclusive hard reactions in free space, the use of nuclear targets to test
and use  color transparency is one of
 its major goals. The $(e,e',p)$ reaction should in particular be studied
in a wide
range of $Q^2$ up to $21 GeV^2$, thus allowing to connect to SLAC data (and
better resolution but similar low  $Q^2$ data from CEBAF) and hopefully clearly
establish this phenomenon in the simplest occurence.
The measurement of the transparency ratio for photo- and
electroproduction of heavy vector mesons, in particular of $\psi$ and
 $\psi'$ will open  a new regime where the mass of the quark enters as
an other scale controlling the formation length of the produced meson.

\section*{Acknowledgments}
 I thank John P. Ralston, Jean-Marc Laget, Jacques Marroncle, Otto Nachtmann,
Eric Voutier, Markus Diehl and Thierry Gousset for many recent discussions on
Color Transparency.  This work has been partially funded through the
European TMR
Contract No FMRX-CT96-0008: Hadronic Physics with High Energy
Electromagnetic Probes.

\end{document}